# On-surface synthesis and collective spin excitations of a triangulene-based nanostar

Jeremy Hieulle+,[b] Silvia Castro+,[a] Niklas Friedrich,[b] Alessio Vegliante,[b] Francisco Romero Lara,[b][c] Sofía Sanz,[d] Dulce Rey,[a] Martina Corso,[c][d] Thomas Frederiksen,[d][e] * Jose Ignacio Pascual,[b][e] * and Diego Peña [a] *

[a]  Dr. S. Castro,[+] Dr. D. Rey, Prof. Dr. D. Peña
     Centro Singular de Investigación en Química Biolóxica e Materiais Moleculares (CiQUS) and Departamento de Química Orgánica
     Universidade de Santiago de Compostela
     15782-Santiago de Compostela (Spain)
     E-mail: diego.pena@usc.es
[b]  Dr. J. Hieulle, N. Friedrich, A. Vegliante, F. R. Lara, Dr. J. I. Pascual
     CIC nanoGUNE-BRTA
     20018 Donostia-San Sebastián (Spain)
[c]  F. R. Lara, Dr. M. Corso
     Centro de Física de Materiales CSIC/UPV-EHU-Materials Physics Center
     20018 Donostia-San Sebastián (Spain)
[d]  S. Sanz, Dr. M. Corso, Dr. T. Frederiksen
     Donostia International Physics Center (DIPC)
     20018 Donostia-San Sebastián (Spain)
[e]  Dr. T. Frederiksen, Dr. J. I. Pascual
     Ikerbasque, Basque Foundation for Science, 48013 Bilbao
     48013 Bilbao (Spain)
[+] These authors contributed equally to this work.

**Abstract:** Triangulene nanographenes are open-shell molecules with predicted high spin state due to the frustration of their conjugated network. Their long-sought synthesis became recently possible over a metal surface. Here, we present a macrocycle formed by six [3]triangulenes, which was obtained by combining the solution synthesis of a dimethylphenyl-anthracene cyclic hexamer and the on-surface cyclodehydrogenation of this precursor over a gold substrate. The resulting triangulene nanostar exhibits a collective spin state generated by the interaction of its 12 unpaired π-electrons along the conjugated lattice, corresponding to the antiferromagnetic ordering of six $S = 1$ sites (one per triangulene unit). Inelastic electron tunneling spectroscopy resolved three spin excitations connecting the singlet ground state with triplet states. The nanostar behaves close to predictions from the Heisenberg model of a $S = 1$ spin ring, representing a unique system to test collective spin modes in cyclic systems.

Open-shell nanographenes have emerged as privilege testing ground to study carbon-based magnetism and to explore potential applications of organic molecules in spintronics and quantum technologies.[1,2] However, the atomically precise preparation of such nanostructures is challenging due to the high reactivity of these graphene derivatives with radical character under ambient conditions. In this sense, on-surface synthesis under ultra-high vacuum (UHV) has become an interesting alternative to fabricate and study the potential functionality of open-shell carbon nanostructures.[3,4] A paramount example is the preparation of [3]triangulene (Figure 1a), a non-Kekulé triangular-shaped polycyclic aromatic hydrocarbon with two unpaired π electrons, which was finally generated and visualized on a surface by means of scanning probe microscopy (SPM),[5] after previous unsuccessful attempts by solution chemistry methods.[6,7] The particular diradical character of [3]triangulene with S = 1, makes it a perfect building block for the design of open-shell carbon-based nanostructures with different ground state spin quantum numbers. In fact, during the last three years, the development of triangulene-based nanographenes has been impressive; not only larger triangulenes, such as [4]-, [5]- and [7]-triangulene,[8-10] but other appealing structures such as a [7]triangulene-ring,[11] an extended version of [3]triangulene,[12] or [3]triangulene dimers[13] were prepared by on-surface synthesis. Even the iconic Clar's goblet, which is actually formed by two fused [3]triangulenes sharing one benzene ring, was generated and characterized on-surface.[14] Interestingly, many of these triangulene-based nanostructures exhibit a characteristic low-energy spectral fingerprint associated with spin excitations induced by inelastic tunneling electrons, which was not observed in isolated triangulene units. Such inelastic features provide unique access to quantify spin interactions in nanographenes and to test current theoretical models for describing collective spin states.

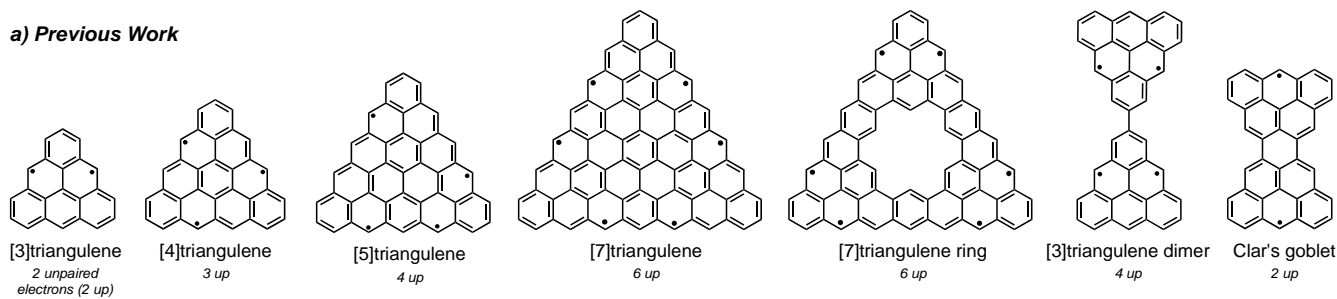

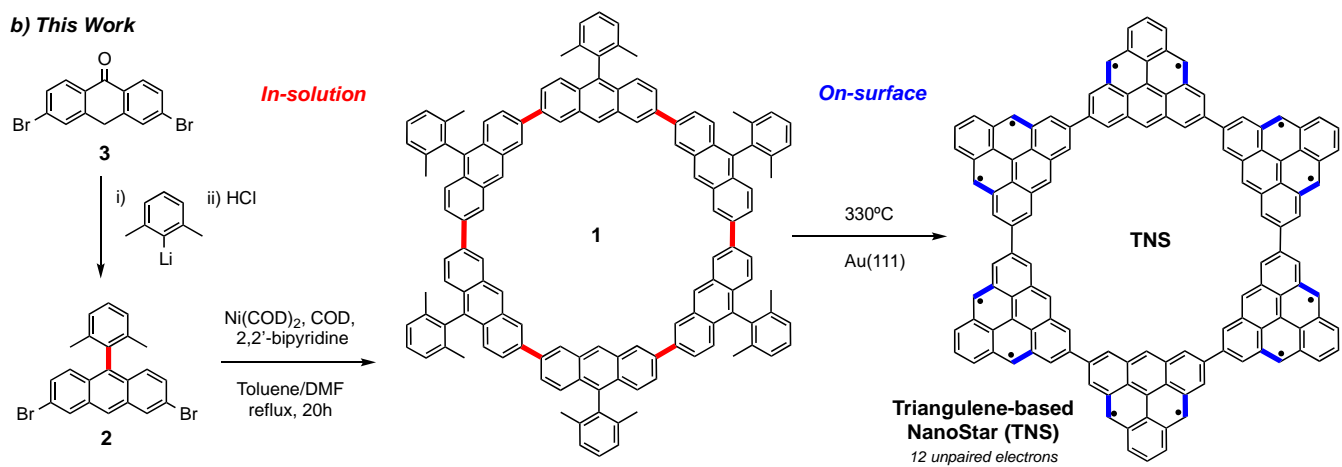

*Figure 1.* a) Previously reported triangulene derivatives. b) Synthesis of triangulene-based nanostar (TNS) developed in this work, by combining in-solution and on-surface chemistry.

In this work, we present a new triangulene derivative consisting in six [3]triangulene moieties arranged in a macrocycle, forming a *triangulene-based nanostar* (**TNS**) with a pore size of 1.2 nm (Figure 1b). Besides the captivating structure of this carbon-based nanosized molecular star, we were particularly intrigued by the spin configuration adopted by the 12 interacting unpaired electrons in the cyclic hexamer over a metal substrate.

We envisioned the preparation of **TNS** by the on-surface cyclodehydrogenation (CDH) of macrocycle **1** (Figure 1b), in particular by the efficient formation of 12 C-C bonds (in blue) and 12 π radicals in one annealing step promoted by Au(111) under UHV conditions. We selected 2,7-dibromo-10-(2,6-dimethylphenyl)-anthracene (**2**) as triangulene building block with the appropriate bromine disposition for the generation of the cyclic hexamer **1** after surface-assisted Ullmann-coupling. Compound **2** was obtained in one step by reaction of dibromoanthrone **3** with (2,6-dimethylphenyl)lithium, followed by acid-promoted dehydration (see supporting information (SI) for details). However, in our hands, the on-surface polymerization of compound **2** resulted in a wide variety of triangulene-based oligomers, as it has been independently reported very recently.[15] Therefore, inspired by the synthesis of *'nano-saturns'* by Toyota and coworkers,[16] we decided to attempt the preparation of macrocycle **1** by means of in-solution coupling of dibromoanthracene **2**. In fact, Ni-catalyzed Yamamoto coupling of compound **2** resulted in a mixture of soluble and insoluble oligomers, which, after purification by filtration and washing procedures, led to the isolation of **TNS** precursor **1** in 10% yield (Figure 1b).

Despite the large molecular mass of macrocycle **1** ($C_{132}H_{96}$, 1680 Da), its vacuum deposition was possible by flash-annealing a silicon wafer loaded with grains of this compound. Figure 2a shows an overview STM image of the Au(111) substrate after deposition. Molecular rings surviving thermal sublimation appear in closed-packed molecular islands, surrounded by polymeric fragments, probably from rings broken during the thermal sublimation (Figure 2a). Individual molecules **1**, extracted from the domains by lateral manipulation with the STM tip, appear composed of six protrusions corresponding to the six dimethylphenyl groups attached to the anthracenes moieties (Figure 2b).

To induce the full planarization of the macrocycle **1** into the targeted **TNS**, we annealed the substrate to 330°C. At this temperature, both the CDH reaction and the formation of π radicals are activated, leading to the removal of 3 H atoms per methyl side group and ring closing through C-C bond formation. However, due to the high temperatures required to fully planarize the whole rings and to their radical character, the CDH competes with additional C-C coupling and bond-breaking reactions, resulting in a few surviving isolated **TNS** and many triangulene oligomers around (Figure 2c and 2d). Bond-resolved scanning tunneling microscopy (BR-STM) images of molecular rings like in Figure 2d, obtained using a CO functionalized STM tip, resolved the triangulene units connected in a closed ring structure (Figure 2e), in agreement with the expected structure of **TNS** (Figure 1b). The planarization reaction involves the formation of twelve six-membered rings, each one at the position of the methyl groups of precursor **1**.

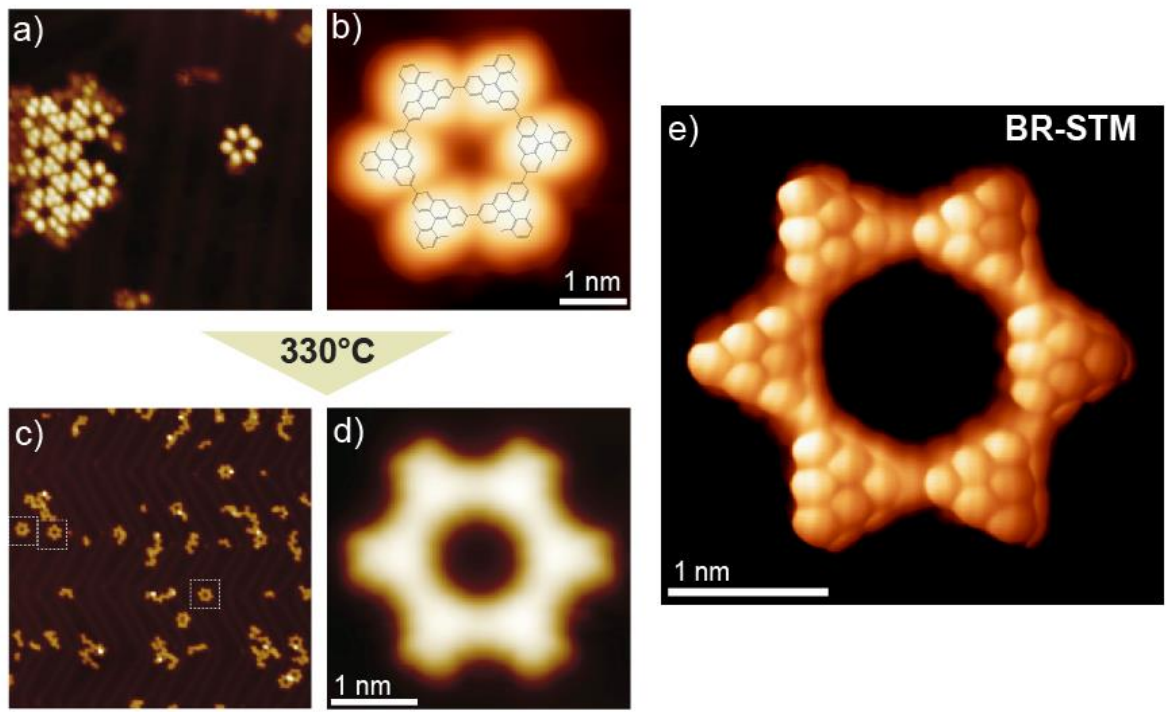

*Figure 2. **On-surface synthesis of triangulene-based nanostar** (TNS). a,b) STM images of precursor 1 as deposited on the Au(111) substrate. Molecular rings appear assembled in packed domains. The isolated molecule 1 in b) was extracted using the STM tip, to prove the covalent nature of the macrocycle. c) STM image of the surface after annealing. TNS and fragments appear mostly planarized. d) STM image of a TNS formed after thermal annealing. e) BR-STM image of a TNS. (a: V = 0.5 V, I = 60 pA; b: V = -0.2 V, I = 40 pA; c: V = 1 V, I = 40 pA; d: V = 1 V, I = 100 pA; e: V = 5 mV)*

The resulting structure accommodates two unpaired π electrons in each triangulene unit, which are expected to interact throughout the conjugated π-lattice and stabilize a collective spin state.[13]

A first inspection of the magnetic interactions in the ring obtained by mean-field Hubbard (MFH) simulations (Figure 3a, see methods in the SI for details) reveals that intra-triangulene parallel magnetic coupling remains strong in the **TNS**, building a total spin $S = 1$ in each triangulene unit, and dominating over a weaker anti-parallel coupling between triangulene sub-units. Such interaction pattern is expected to lead to a global spin singlet ground state in the **TNS**. Interestingly, antiferromagnetic (AFM) rings of integer spins have been widely studied for demonstrating the Haldane conjecture,[17,18] stating that the spin excitation spectrum of an infinite Heisenberg $S = 1$ AFM chain remains gapped, with an energy value close to 0.4 *J* (*J* being the Heisenberg exchange constant). Such a gapped spectrum is an anomalous behavior of integer AFM spin chains, which contrasts with the gapless spin excitation of similar chains made of half-integer spins.[18]

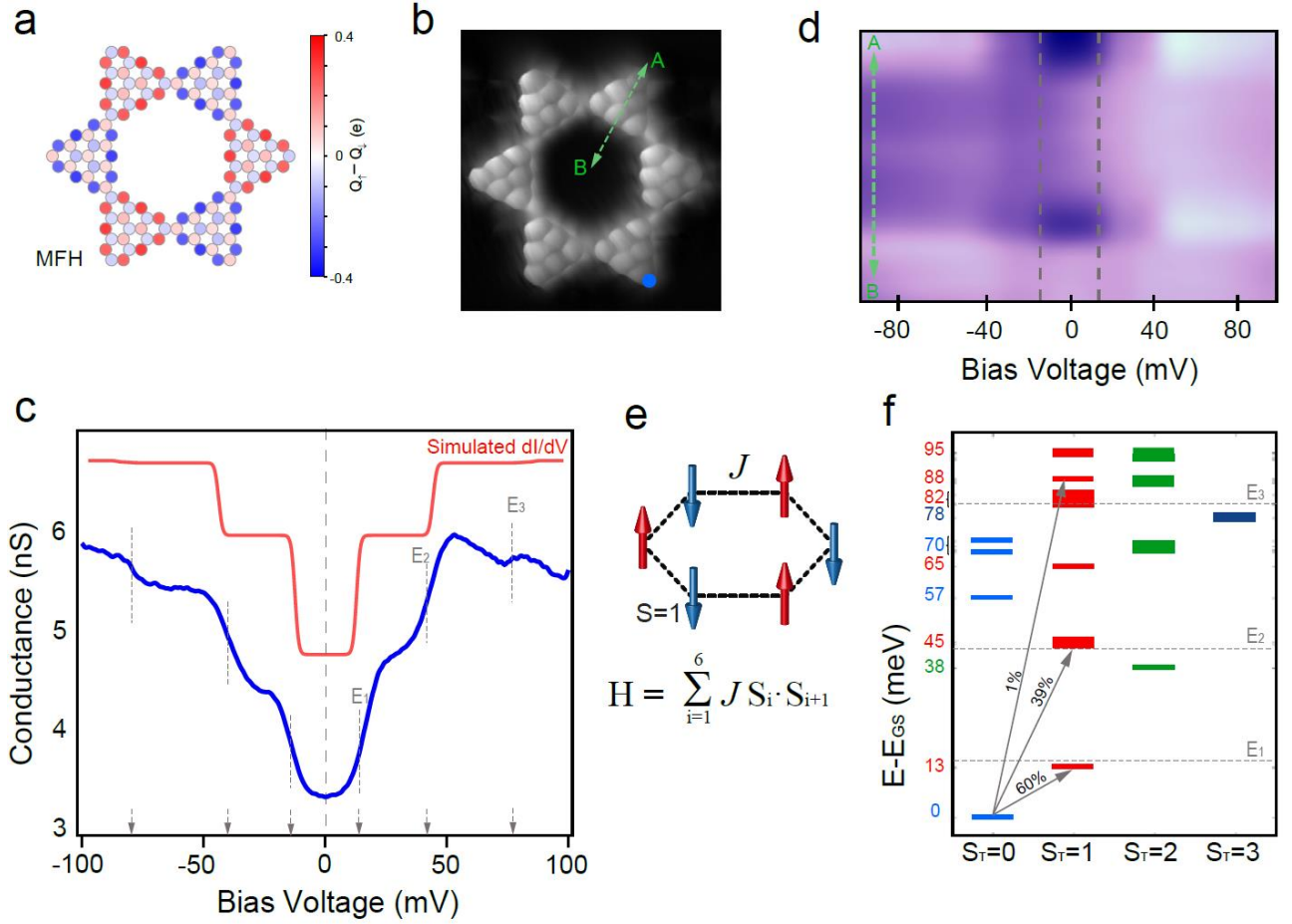

*Figure 3. Collective spin excitations in a triangulene-based nanostar.* a) Spin density map of a TNS obtained from mean-field Hubbard simulations (on-site Coulomb potential U = 3.5 eV). b) BR-STM image indicating the TNS site where spectra in c) and line spectra in d) were taken. c) dI/dV spectrum on a triangulene unit of a TNS, showing inelastic steps at E1 = ±14 mV, E2 = ±42 mV and E3 = ±80 mV. The red plot shows simulated spectral function after the results of a Heisenberg model with exchange constant J = 18 meV. d) dI/dV spectral line measured across a single triangulene unit. The excitation steps appear

To experimentally determine the spin configuration of the **TNS**, we probed their excitation spectrum by Inelastic Electron Tunneling Spectroscopy (IETS).[19,20] Specifically, we measured differential conductance (dI/dV) vs. bias spectra over the sides of triangulene units of a **TNS** and found bias-symmetric stepped features (Figure 3c), which are a characteristic fingerprint of spin excitation induced by inelastic electron tunneling.[13,14,21] All triangulene units show spectra with the same shape (see additional data in SI), composed of three steps of different height at $E_1 = \pm14(1)$ mV, $E_2 = \pm42(2)$ mV and $E_3 = \pm80(2)$ mV. The first two dI/dV steps have similar height (each amounting to a relative increase of conductance of ~35%), while the third one appears with small spectral weight (>7%). As we show in the stacked dI/dV plot of Figure 3d, the inelastic signal appears localized over the edges of the triangulene units, in agreement of the localization sites expected for their π-radical states.

The presence of three spin excitation steps in the dI/dV spectra is a remarkable feature that contrasts with the single-stepped spectra measured in triangulene dimers[13] and other graphene nanostructures[14,21–23]. Due to the $\Delta S = 0, \pm 1$ selection rule imposed by conservation of angular momentum,[19] single IETS steps were attributed to singlet-triplet transitions induced by the tunneling electrons. The multi-step spectrum found for the **TNS** unveils a complex excitation pattern of collective spin states in the triangulene ring, accessible by inelastic tunneling electrons.

To explain the origin of the inelastic steps, we modeled the coupled spin system with an isotropic Heisenberg Hamiltonian describing the AFM exchange interaction of six S = 1 spins in a closed ring, as in Figure 3e (theoretical details in SI). The many-body ground state of the TNS, obtained by numerically exact diagonalization of the spin Hamiltonian (see methods in SI), has a total spin $S_T = 0$ (singlet), and consists of the superposition of the classical solutions of antiferromagnetically coupled triangulene segments (i.e. ground state from the MFH model,[24] in Figure 3a), and other combinations of the triangulene spins with global spin $S_T = 0$. Similarly, the set of excited states of the TNS are collective spin modes with total spin amounting from $S_T = 0$ to $S_T = 6$, which can be described as a superposition of triangulene S = 1 multiplet states.

To explain the spin excitation spectrum, we assume that steps correspond to transitions with $\Delta S = \pm 1$[25] and adjust the Heisenberg exchange constant $J$ to fit the lowest allowed excitation with the first dI/dV step (centered at ±14 mV). The best spectral match occurs using $J = 18$ meV. This value is larger than the exchange constant derived from the Heisenberg model in triangulene dimers.[13] Figure 3f shows the computed excitations with energy up to 100 meV, classified according to their global spin (up to $S_T = 3$). Several singlet and triplet excited states appear within the energy window of the IETS spectrum, including triplet states grouped around excitation energy values (13 meV, 45 meV and 85 meV) that agree with the three dI/dV steps in our experimental spectra. However, not all transitions are equally accessible by tunneling electrons. To account for the cross-section of each spin transition and explain their different dI/dV step height, we calculated the correlation of collective **TNS** spin states with tunneling electrons following the expression by Ternes.[26] This quantity reflects the scattering probability of a tunnel electron with every collective spin eigenstate and, thus, it is related to the inelastic signal and, hence, to the step height. The first excitation at 13 meV has the larger spectral weight (~60%, as shown by arrows in Figure 3f), while several states bundled at around $E_2 = 45$ meV contribute to most of the remaining weight (39%). Higher states are seemingly weaker in terms of spectral weight and produce a faint step in spectrum. A simulated spin excitation spectrum shown in Figure 3c (red plot) reproduces with good agreement the position and relative intensity of the three steps observed in the experimental plot. There are, however, small differences in step position and height that can be accounted for by including a more precise description of spin interactions. For example, including a finite intra-triangulene coupling can correctly reproduce the equal height of first and second steps (see analysis for twelve 1/2 spins in the SI). Other effects such as a renormalization of spin interactions in the **TNS** induced by the metal substrate,[27] or higher order correction to the Heisenberg coupling between spins,[15,18] can be additionally incorporated into the model to quantitatively reproduce the experimental spectra.

In summary, we have demonstrated the successful fabrication of a macrocycle formed by the covalent attachment of six [3]triangulenes, through a combination of solution synthesis coupling and on-surface planarization. The design-driven strategy of precursor molecules is thus a possible method to produce monodispersed ensembles of complex open-shell nanostructures. For the presented case, the triangulene cyclic hexamer accumulated 12 unpaired electrons. Tunnel spectroscopy revealed that the unique spin states expected for such interacting systems survive on a metal surface. We detected three collective spin excitations from the many-body singlet ground state to triplet states of the spin ring. These findings proved that the spin states of the nanostar are compatible with an antiferromagnetic S = 1 spin ring.

**Acknowledgements**

We acknowledge funding from *Agencia Estatal de Investigación* (PID2019-107338RB, FIS2017-83780-P, MDM-2016-0618), from the *Xunta de Galicia (Centro singular de investigación de Galicia*, accreditation 2019-2022, ED431G 2019/03), the University of the Basque Country (Grant IT1246-19), the European Union H2020 program (FET Open project SPRING #863098), the European Regional Development Fund and the Basque *Departamento de Educación* for the PhD scholarship no. PRE_2020_2_0049 (S.S.) The authors thank D. Pérez and E. Guitián for fruitful discussions.

# On-surface synthesis and collective spin excitations of a triangulene-based nanostar

Jeremy Hieulle[+],[b] Silvia Castro[+],[a] Niklas Friedrich,[b] Alessio Vegliante,[b] Francisco Romero Lara,[b][c] Sofía Sanz,[d] Dulce Rey,[a] Martina Corso,[c][d] Thomas Frederiksen,[d][e] * Jose Ignacio Pascual,[b][e] * and Diego Peña [a] *

---

[a]  Dr. S. Castro,[+] Dr. D. Rey, Prof. Dr. D. Peña
    Centro Singular de Investigación en Química Biolóxica e Materiais Moleculares (CiQUS) and Departamento de Química Orgánica
    Universidade de Santiago de Compostela
    15782-Santiago de Compostela (Spain)
    E-mail: diego.pena@usc.es
[b]  Dr. J. Hieulle, N. Friedrich, A. Vegliante, F. R. Lara, Dr. J. I. Pascual
    CIC nanoGUNE-BRTA
    20018 Donostia-San Sebastián (Spain)
[c]  F. R. Lara, Dr. M. Corso
    Centro de Física de Materiales CSIC/UPV-EHU-Materials Physics Center
    20018 Donostia-San Sebastián (Spain)
[d]  S. Sanz, Dr. M. Corso, Dr. T. Frederiksen
    Donostia International Physics Center (DIPC)
    20018 Donostia-San Sebastián (Spain)
[e]  Dr. T. Frederiksen, Dr. J. I. Pascual
    Ikerbasque, Basque Foundation for Science, 48013 Bilbao
    48013 Bilbao (Spain)
[+] These authors contributed equally to this work.

## *Supporting Information*

### *Table of Contents*

# 1. In-solution synthesis and characterization

## 1.1 General details

All reactions were carried out under argon using oven-dried glassware. Et$_2$O, toluene and DMF were dried using a MBraun SPS-800 Solvent Purification System. Other commercial reagents were purchased from ABCR GmbH, Sigma -Aldrich or Acros Organics, and were used without further purification. Deuterated solvents were purchased from Acros Organics. Thin layer chromatography was performed on Merck silica gel 60 F 254 and chromatograms were visualized with UV light (254 and 365 nm) and/or stained with Hanessian's stain. Column chromatography was performed on Merck silica gel 60 (ASTM 230-400 mesh). $^1$H and $^{13}$C NMR spectra were recorded at 300 and 75 MHz (Varian Mercury-300 instrument) or 500 and 125 MHz (Varian Inova 500 or Bruker 500) respectively. Low-resolution mass spectra (EI) were obtained at 70 eV on a HP5988A instrument, while high-resolution mass spectra (HRMS) were obtained on a Micromass Autospec spectrometer. APCI high resolution mass spectra were obtained on a Bruker Microtof instrument. MALDI mass spectra were obtained on a Ultraflex III TOF/TOF Bruker instrument. Chemical shifts are reported in ppm and referenced to residual solvent. Coupling constants (*J*) are reported in Hertz (Hz). Standard abbreviations indicating multiplicity were used as follows: m = multiplet, quint. = quintet, q = quartet, t = triplet, d = doublet, s = singlet, b = broad.

## 1.2 Synthesis of precursors

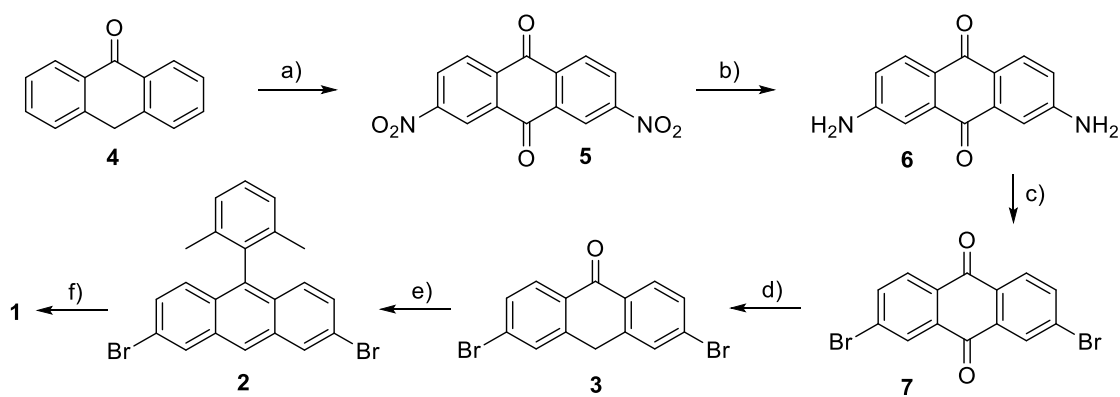

**Scheme S1.** Synthesis of compound **1**. Compound **3** was prepared from commercially available compound **4** following reported procedures.[1] a) 1. HNO$_3$, 2. AcOH, 1 week, 35%; b) Na$_2$S, NaOH, EtOH/H$_2$O, 1 day, 90% c) 1. CuBr, *t*-BuNO$_2$, 2. HCl, CH$_3$CN, 45%; d) Al powder, H$_2$SO$_4$, 90%; e) 1. 1,3-dimethyl-2-iodobenzene, *t*-BuLi, Et$_2$O, Toluene, 2. HCl, 35 %; f) Ni(COD)$_2$, COD, 2,2'-bipyridine, toluene/DMF, reflux, 20h, 10%.

**Synthesis of 2,7-dibromo-10-(2,6-dimethylphenyl)-anthracene (2)**

In a dry Schlenk flask, anhydrous Et$_2$O (2 mL) and 1,3-dimethyl-2-iodobenzene (115 µL, 0.810 mmol) were introduced under Ar atmosphere and cooled to -78 ºC. Then, *t*-BuLi (1.0 mL of 1.6 M hexane solution, 1.6 mmol) was added dropwise. After stirring at -78 ºC for 1h, anhydrous toluene (2 mL) was added and the solution was warmed up to 60 ºC. Then, a solution of 3,6-dibromo-9-anthrone (190 mg, 0.540 mmol) in toluene (6 mL) was added via cannula. The mixture was stirred at 60 ºC for 1h. After this time, conc. HCl (2.0 mL) was added dropwise and the reaction mixture was diluted with hexane and transferred to a separation funnel. The organic layer was separated and washed with water. The aqueous layer was extracted with hexane. The combined organic layers were dried over anhydrous MgSO$_4$ and the solvent was removed under vacuum. The crude was purified by flash column chromatography (SiO$_2$, Hexane/DCM (3:1)) affording **2** in a 35 % yield. $^1$H NMR (500 MHz, CDCl$_3$, Figure S1): δ = 8.34 (s, 1H), 8.25 (d, *J* = 1.6 Hz, 2H), 7.44 – 7.24 (m, 7H), 1.70 (s, 6H). $^{13}$C NMR (125 MHz, CDCl$_3$, Figure S2-3): δ = 137.95 (C), 137.41 (C), 136.68 (C), 133.36 (C), 130.61 (CH), 129.91 (CH), 128.59 (CH), 128.45 (C), 128.30 (CH), 128.06 (CH), 124.85 (CH), 120.67 (CH). APCI-MS (Figure S4): m/z calcd. for C$_{22}$H$_{16}$Br$_2$ [M+H$^+$]: 440.9672, found: 440.9675.

**Synthesis of macrocycle 1**

In a dry Schlenk flask, a mixture of Ni(COD)$_2$ (118 mg, 0.429 mmol), COD (53 µL, 0.429 mmol) and 2,2'-bipyridine (67 mg, 0.429 mmol) in anhydrous toluene (360 µL) and DMF (360 µL) was prepared and degassed by bubbling Ar for 10 min. Then, the mixture was heated at 80 ºC for 30 min. A solution of compound **2** (80 mg, 0.182 mmol) in toluene (1.28 mL) was added via syringe and the temperature was rise up to 110 ºC. The mixture was stirred at this temperature overnight. After cooling to room temperature, HCl (2.0 mL, 1.0 M) was added and the mixture was stirred for 1h. The solid fraction was separated by filtration and washed with MeOH, hexane and DCM. Then, the solid was washed with THF in a Soxhlet extractor for 1 day, to give a yellow insoluble solid. MALDI-TOF mass spectrometry confirmed the molecular formula of compound **1** (10 % yield). Characterization by $^1$H NMR was not possible due to the extreme low solubility of this compound. MALDI-TOF MS (Figure S5-6): m/z calcd. for C$_{132}$H$_{96}$ [M+]: 1681.8 found: 1681.9.

The filtrate was separately analyzed by mass spectrometry showing a mixture of different oligomers derived from precursor **2**. After separation by recycling gel permeation chromatography, a cyclic heptamer was isolated and detected by MALDI (Figure S6-7).

*1.3 Spectroscopic Data*

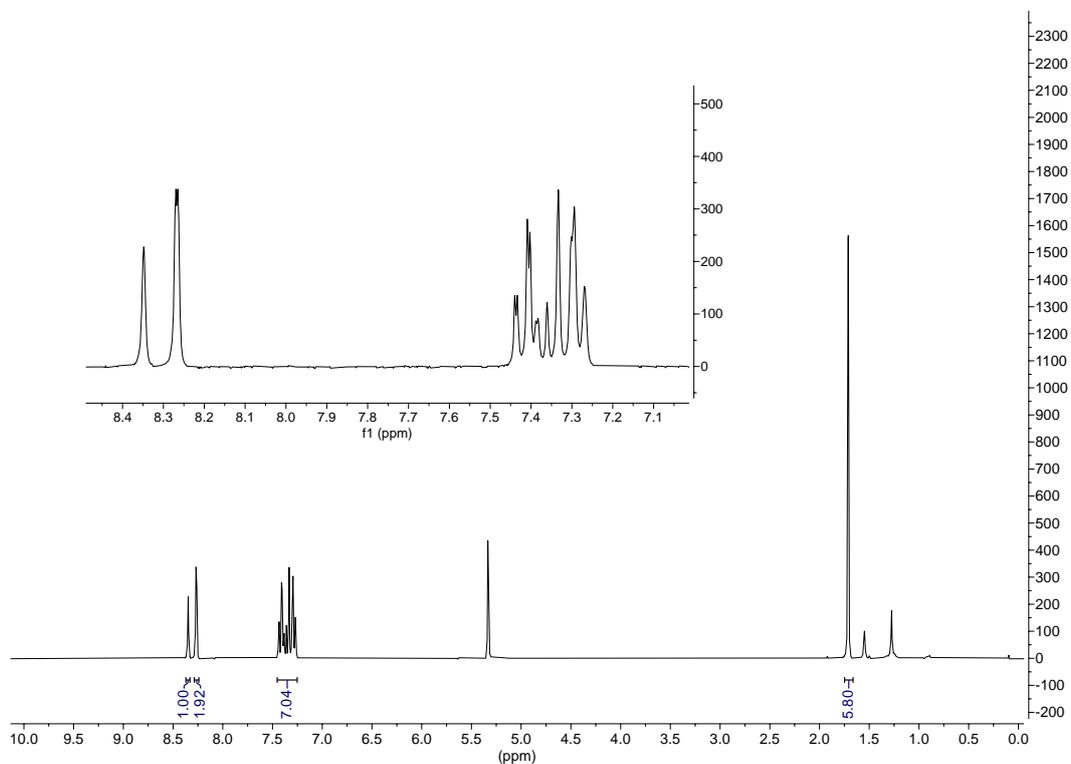

**Figure S1.** $^1$H NMR (500 MHz, CDCl$_3$) of compound **2**.

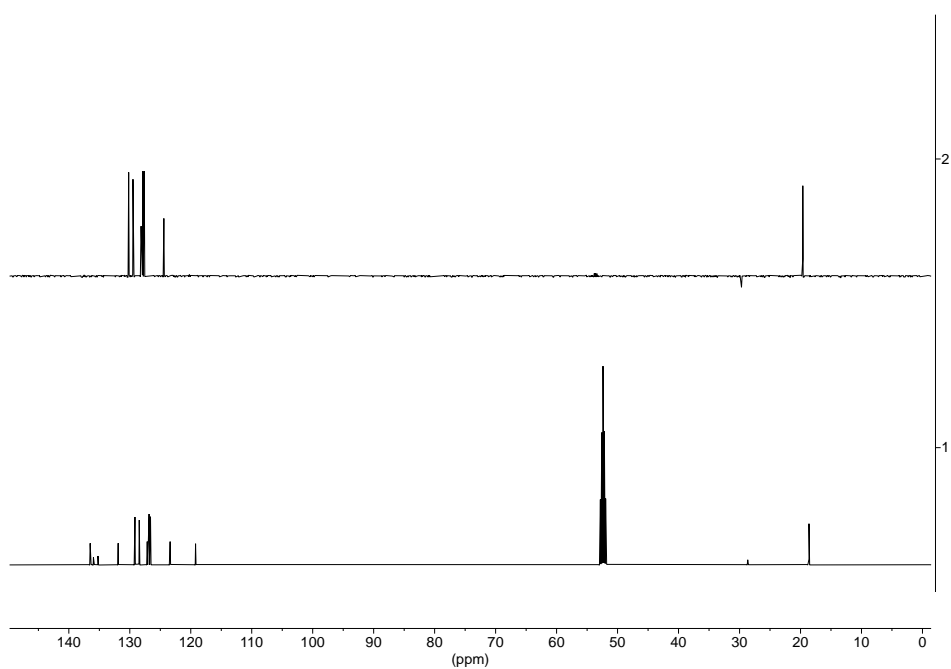

**Figure S2.** $^{13}$C (bottom) and DEPT 135 (top) NMR (125 MHz, CDCl$_3$) of compound **2**.

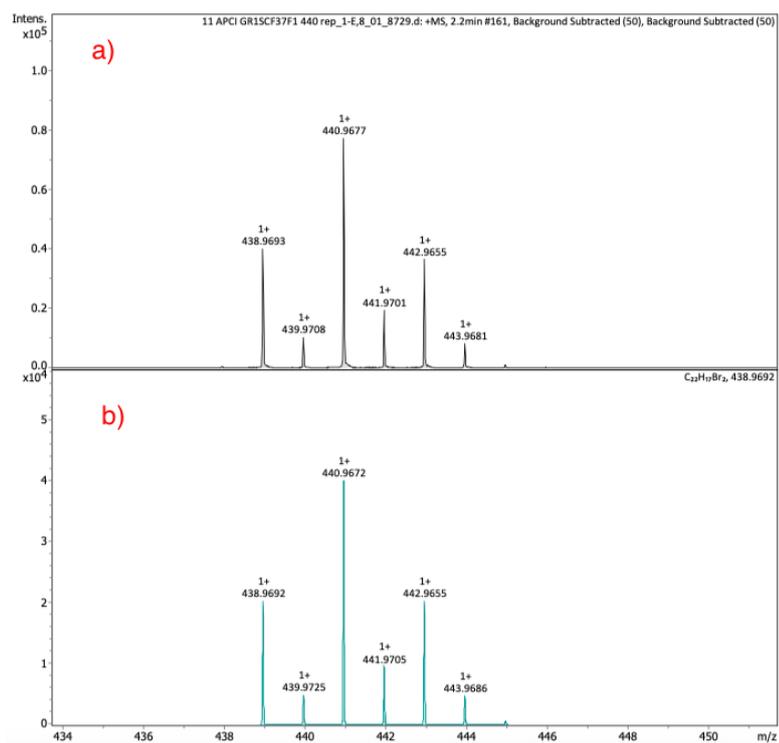

**Figure S3.** a) HRMS APCI of compound **2**; b) Simulated mass spectrum of compound **2**.

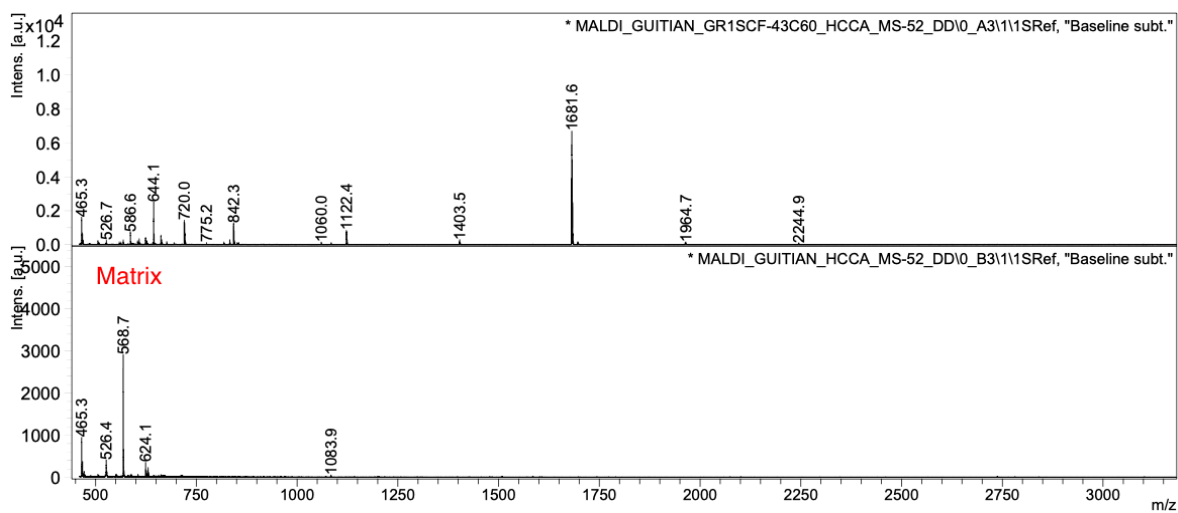

**Figure S4.** MALDI-TOF MS of compound **1** using α-cyano-4-hydroxycinnamic acid as matrix (HCCA).

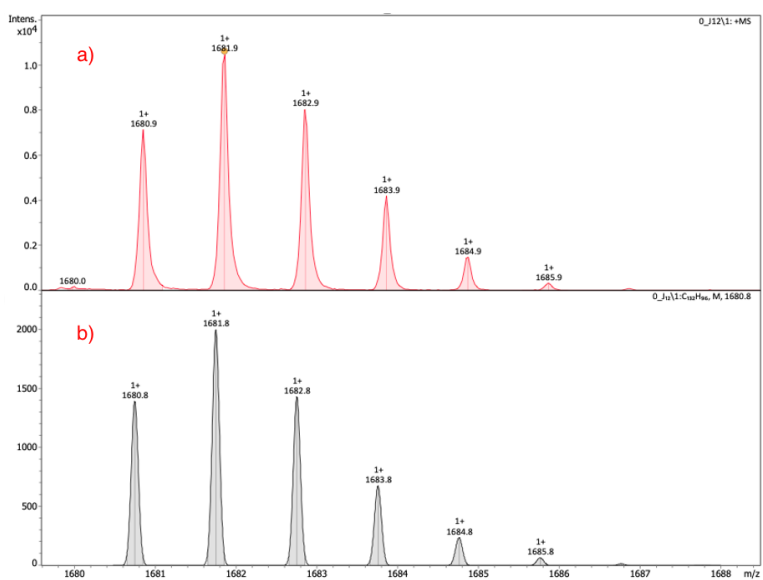

**Figure S5.** Zoom of the MALDI-TOF MS spectrum of compound **1**: a) Experimental; b) Simulation.

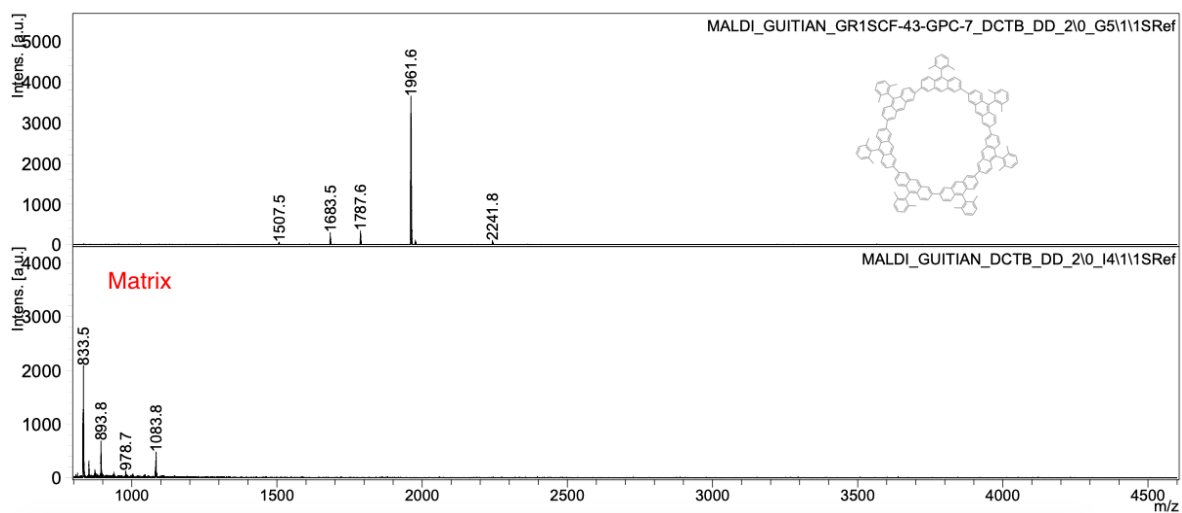

**Figure S6.** MALDI-TOF MS of a cyclic heptamer, using α-cyano-4-hydroxycinnamic acid as matrix (HCCA).

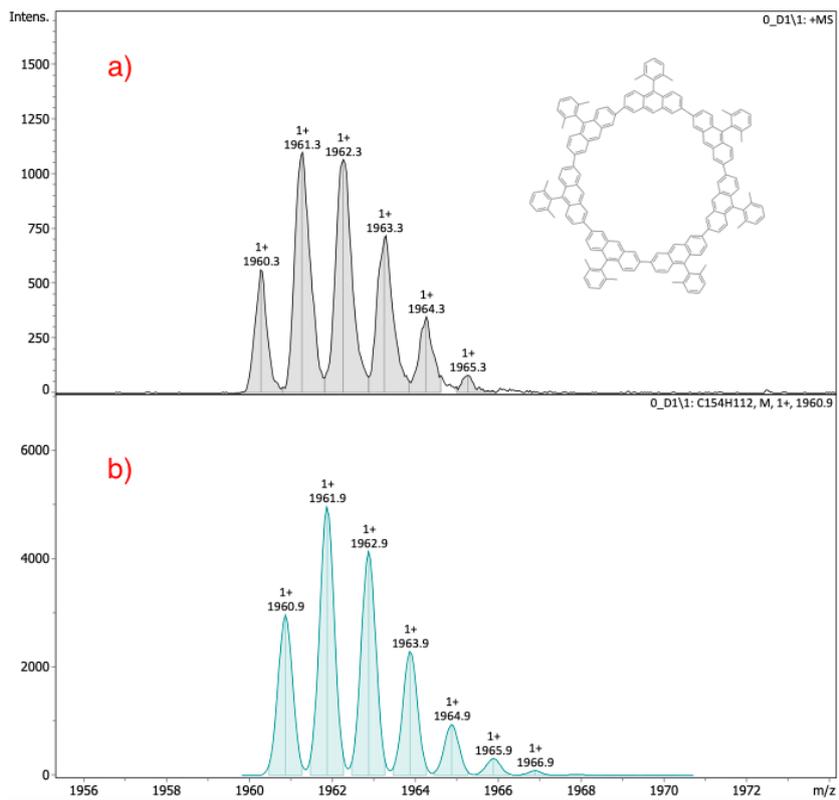

**Figure S7.** Zoom of the MALDI-TOF MS spectrum a cyclic heptamer: a) Experimental; b) Simulation.

## 2. On-surface synthesis and characterization

*2.1 Experimental details*

The experiments were performed in an ultra-high vacuum (UHV) low temperature scanning tunneling microscope, at a base temperature of 5 K. The Au(111) surface of a gold single crystal was cleaned by several cycles of sputtering with Ne gas and annealing at 460°C. A few grains of precursor 1 in Fig. 1 were loaded on a Si wafer that can be heat up to 600°C in a few seconds, and inserted into the UHV system for its flash deposition on to the clean Au(111) substrate. Due to the high molecular mass of molecular precursor **1**, the use of flash-evaporator is essential with respect to usual Knudsen cell. The triangulene-based nanostar (**TNS**) were prepared by annealing the Au(111) substrate to 330°C for about 10 minutes for inducing the on-surface cyclodehydrogenation reaction of the deposited molecular precursor **1**. The sample was characterized by STM and STS. Bond-resolved STM data were obtained by functionalizing the tip with a CO molecule and recording constant height images in the Pauli repulsion regime, for a typical voltage and current of 5 mV and 500 pA, respectively. Scanning tunneling spectra were acquired with a lock-in amplifier to obtain the derivative of the current versus voltage curve (dI/dV), by applying a small modulation voltage $V_{ac}$= 2 mV r.m.s.

*2.2 Theoretical details*

**Mean-field Hubbard simulations:** Mean-field Hubbard (MFH) calculations were performed using the Python code we developed and documented elsewhere.[2, 3] The MFH Hamiltonian for the carbon $p_z$ orbitals includes first- to third-nearest neighbour hopping terms as well as on-site Coulomb repulsion $U$ within the mean-field approximation. The expectation value of the spin-polarized density operator $n_{i\sigma} = c_{i\sigma}^{\dagger} c_{i\sigma}$ is calculated from the self-consistent solution to the eigenvalue problem of the MFH Hamiltonian. We used the parameters $t_1 = 2.7 eV, t_2 = 0.2 eV, t_3 = 0.18 eV, U = 3.5 eV$ for the presented simulation.

**Isotropic Heisenberg Spin Hamiltonians:** The differential conductance spectra were modelled using the code developed by Markus Ternes,[4] as well as our own implementation in Python of Heisenberg spin Hamiltonians. The modelled system consists of six $S = 1$ spin sites that are coupled antiferromagnetically (AFM) with the exchange coupling $J = 18$ meV to each two other spins such that the system forms a closed ring:

$$\hat{H} = J \sum_{i=1}^{6} \hat{\mathbf{S}}^{(i)} \cdot \hat{\mathbf{S}}^{(i+1)}, \qquad \hat{\mathbf{S}}^{(i)} = \hat{\mathbf{S}}^{(i+6)}$$

Where $\hat{\mathbf{S}}^{(i)} = (\hat{S}_x^{(i)}, \hat{S}_y^{(i)}, \hat{S}_z^{(i)})$ is the quantum mechanical spin-1 operator for site $i$. We used an effective temperature of $T_{eff} = 5K$ to account for the finite temperature in experiment. We considered the complete set of $3^6 = 729$ different linear combinations of $S_z$-projections available for the product state of six $S = 1$ spins. Numerically exact diagonalization of the spin Hamiltonian provides the eigenstates $|n\rangle$ and eigenenergies $E_n$. The relative transition rates for each transition, which scales the height of the conductance steps in the experiment, are obtained via the expression:

$$\left|M_{mn}^{(i)}\right|^2 = \frac{1}{2}\left|\langle m|\hat{S}_+^{(i)}|n\rangle\right|^2 + \frac{1}{2}\left|\langle m|\hat{S}_-^{(i)}|n\rangle\right|^2 + \left|\langle m|\hat{S}_z^{(i)}|n\rangle\right|^2$$

where $(\hat{S}_+^{(i)}, \hat{S}_-^{(i)}, \hat{S}_z^{(i)})$ are the lowering, raising, and $z$-projection spin-1 operators corresponding to site $i$.

### *2.3 Comparison of differential conductance spectra on the different triangulene units*

Figure S8 compares dI/dV-spectra obtained at equivalent positions on each of the six triangulene units forming a **TNS**. The spectra show only minor differences in every triangulene subunit.

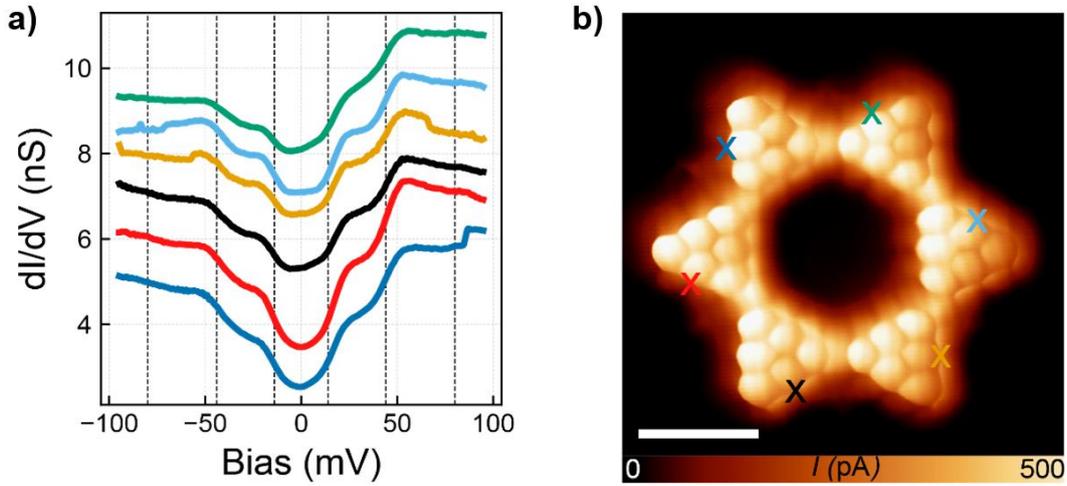

**Figure S8.** a) Differential conductance spectra on the six triangulene units of a **TNS**. Spectra are taken at the position indicated in b). Spectra are offset by 1 nS for clarity. See main text for details of the lock-in amplifier. b) Constant height current image of a **TNS**. $V = 5mV$.

## 2.4 Transition rates of the active excitations of a spin 1 triangulene ring

Here, we present results from the solutions of the Heisenberg spin Hamiltonian, indicating the eigen states with energy below 100 meV, their degeneracy (including $S_z$ multiplets), and the transition rates into the final state. As we show here, the excitation spectrum for scattering with tunneling electrons is dominated by three collective spin excitations.

| ENERGY VALUES (meV) | DEGENERACY | FINAL STATE | TRANSITION RATE | TRANSITION RATE (% norm) | %Excitation |
|---|---|---|---|---|---|
| 0 | 1 | 1 | 0,40960853 | 20,003 | |
| 12,97 | 3 | 2 | 0,40960854 | 20,003 | 60,009 |
| 38,80 | 5 | 3 | 0,40960855 | 20,003 | |
| 44,92 | 6 | 9 | 0,17370307 | 8,483 | |
| 45,59 | 6 | 10 | 0,01507207 | 0,736 | |
| 57,02 | 1 | 11 | 0,03114157 | 1,521 | |
| 65,11 | 3 | 12 | 0,15816743 | 7,724 | |
| 68,75 | 1 | 13 | 0,1853627 | 9,052 | |
| 69,46 | 10 | 14 | 0,02403847 | 1,174 | |
| 70,41 | 10 | 15 | 0,02500253 | 1,221 | 39,272 |
| 71,84 | 2 | 16 | 0,03546538 | 1,732 | |
| 77,73 | 7 | 17 | 0,06588653 | 3,218 | |
| 81,42 | 6 | 18 | 0,02856793 | 1,395 | |
| 83,77 | 6 | 19 | 0,02181164 | 1,065 | |
| 87,04 | 10 | 20 | 0,0399678 | 1,952 | |
| 87,85 | 3 | 77 | 0,00490359 | 0,239 | |
| 93,04 | 5 | 78 | 0,00490359 | 0,239 | 0,718 |
| 93,66 | 5 | 79 | 0,00490359 | 0,239 | |
| 94,62 | 6 | **SITE 0 and 3** | | **100,000** | |

## 2.5 Comparison with a Heisenberg ring of twelve S = 1/2 spins

The Heisenberg model utilized in the main manuscript considers that two-spins per triangulene couple with infinite energy to form a S = 1 state. We explored the effect of a finite intra-triangulene ferromagnetic exchange coupling on the Heisenberg spin excitation. In this case we employ a 12-site spin-1/2 Heisenberg ring:

$$\hat{H} = 4J' \sum_{i=1}^{6} \hat{\mathbf{S}}^{(2i-1)} \cdot \hat{\mathbf{S}}^{(2i)} + 4J \sum_{i=1}^{6} \hat{\mathbf{S}}^{(2i)} \cdot \hat{\mathbf{S}}^{(2i+1)}, \qquad \hat{\mathbf{S}}^{(i)} = \hat{\mathbf{S}}^{(i+12)}$$

where $J > 0$ and $J' < 0$ describe the (antiparallel) inter- and (parallel) intra-triangulene exchange couplings, respectively. Now $\hat{\mathbf{S}}^{(i)} = (\hat{S}_x^{(i)}, \hat{S}_y^{(i)}, \hat{S}_z^{(i)})$ is the spin-1/2 operator (two spin-1/2 sites included per triangulene). For comparison between the two case models a factor 4 is conveniently included (arising from normalization of the spin operators).

The comparison between the two models is shown in Figure S9 for fixed $J = 18$ meV and varying ratios of the intra-triangulene ferromagnetic (FM) exchange coupling $J'$ (blue and red curves in Figure S9). Indeed, when $|J'| \gg |J|$ the transition energies of the two models coincide. The step heights are somewhat different between the two models. In fact, by including a large but finite value for $J'$ (e.g. $-J'/J=10$ or 100), we observe that the step heights for the first and for the second excitation become similar in magnitude, as we find in the experiment and in contrast to the ~60/40 ratio we obtain for a S = 1 model (see Figure 3f in the manuscript). Thus, a finite intra-triangulene coupling between the two unpaired electrons simulate better the experimental results.

We also considered the effect of a small $J'$ in our interpretation of collective spin states. As $J'$ is reduced, one observes an upward shift of the excitation thresholds, with the first threshold shifting the most. If we fix the ratio $-J'/J=5$, we can fit the first experimental threshold energy by lowering the AFM coupling to $J=14$ meV (green dashed curve in Figure S9). However, then the second threshold does not match very well. This suggests that the assumption $|J'| \gg |J|$ is still reasonable to describe our experiments.

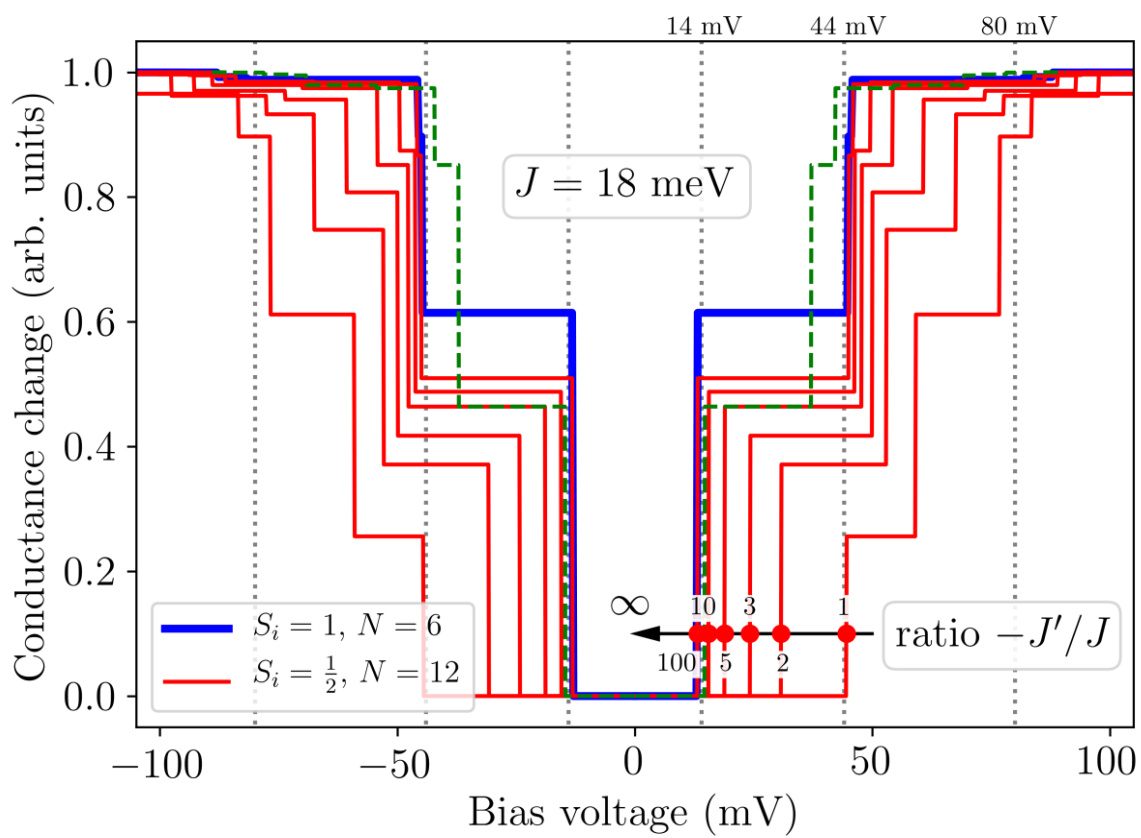

**Figure S9**: Calculated conductance changes for different Heisenberg spin ring models. (blue) ring of six spin-1 sites, (red) ring of 12 spin-1/2 sites with varying intra- and inter-triangulene exchange couplings. (green) ring of 12 spin-1/2 sites with $J=14$ meV and ratio $-J'/J = 5$.